\documentclass[review]{elsarticle}

\usepackage{lineno,hyperref,latexsym}

\journal{Computational Biology and Chemistry}

\begin{document}

\begin{frontmatter}

\title{A model for the clustered distribution of SNPs in the human genome}

\author{Chang-Yong Lee\fnref{myfootnote}}
\address{The Department of Industrial and Systems Engineering, Kongju
  National University, Cheonan, 330-717, South Korea}
\fntext[myfootnote]{clee@kongju.ac.kr}




\begin{abstract}
Motivated by a non-random but clustered distribution of SNPs, we
introduce a phenomenological model to account for the clustering
properties of SNPs in the human genome. The phenomenological model is
based on a preferential mutation to the closer proximity of existing SNPs. 
With the Hapmap SNP data, we empirically demonstrate
that the preferential  model is better for illustrating the clustered
distribution of SNPs than the random model. 
Moreover, the model is applicable not only to autosomes
but also to the X chromosome, although the X chromosome has different
characteristics from autosomes. 
The analysis of the estimated parameters in the model can explain the
pronounced population structure and the low genetic diversity of the X
chromosome. In addition, correlation between the parameters
reveals the population-wise difference of the mutation probability. 
These results support the mutational non-independence
hypothesis against random mutation.
\end{abstract}


\end{frontmatter}


\section{INTRODUCTION}
The most common type of genetic variants in the human genome is the single
nucleotide polymorphism (SNP), which, as a result of mutation, has a
difference in a single nucleotide within a population of
samples~\cite{barr}.   
SNP data, together with gene expression and other biological
information, are an important resource to answer various biological
questions regarding the genetic variation, such as the mutational
pattern of the genome, the phylogenetic classification, and the association 
with phenotype data.   

In recent years, as the cost of genotyping has dropped dramatically due mainly
to the advance in the genotyping technology~\cite{metzker,array},
much effort has been put into the identification of SNPs in the human
genome~\cite{hap,100}.  
Notably, the International HapMap project~\cite{hap} (hereafter,
Hapmap) is an international effort to identify the genetic variation in
the human genome to develop a haplotype map. Although Hapmap includes some
datasets on the copy number variation, SNP data are the main resource not
only for understanding and characterizing the differences in genome
structure but for association studies with diseases and/or
environmental factors.   

It has been known that SNPs in the human genome are not distributed
randomly but clustered across the genome~\cite{amos,tena,kobo,hell,lind}.
This clustering property suggests that mutations tend to occur not
randomly but preferentially to the proximity of existing mutations.
In addition to the interpretation of the clustering as the reflection
of mutational hotspots~\cite{rogo}, clustered SNPs can emerge in
various ways. Natural and balancing selections can 
modulate local variability and tend to create regions of increased
variability that results in non-randomness~\cite{bubb}. A high
variance of genes within a population of samples in the time to the
most recent common ancestor causes different recombination rates in a
chromosome~\cite{erik}.
It was also proposed that microsatellites can also generate mutational
biases in their flanking regions by expansion and erosion from the
perspective of microsatellite evolution~\cite{vowles,webster,varela}.
Clustered SNPs can also arise from ascertainment biases in the SNP
discovery process~\cite{kuh}. Examples include the SNP identification
based on maximally dissimilar sequences, the usage of not enough
samples, and finding all possible SNPs not on a whole genome but on
a given region of a chromosome. 

However, when SNP clusters are found throughout a whole genome with a
large number of samples from different global populations, it is
unlikely that the observed clusters would be due to ascertainment biases. 
Thus, as pointed out in Ref.~\cite{amos}, the majority of SNP markers
along a whole genome should reflect the underlying mutation pattern. 
In this respect, a non-random
mutation process was proposed and tested 
against the random mutation by generating a semi-realistic population
of chromosomes from stochastic computer simulations that implements
the concept of `the sphere of influence'~\cite{amos}.    
As millions of SNPs on a whole genome are now available in public
domains, the mutation pattern can be systematically investigated.

In this paper, we propose a probabilistic model for the clustered
distribution of SNPs. The proposed model assumes non-independent
mutations in which subsequent mutations occur not randomly but
preferentially to near mutated sites.  
Within the model, SNP clusters could form mainly through a
non-negligible tendency of the mutation process in the closer proximity  
of existing SNPs. The proposed model was tested against Hapmap SNP
data and the proposed model was confirmed as suitable
to explain empirical SNP distributions of the human genome.
We also tested the proposed model against the random mutation model
in which all mutations occur independently, and we confirm that the
proposed model explains the distribution more appropriately than
the random model.  

As the X chromosome is a haploid in males, its SNP distribution may
have characteristics different from the distributions of the autosomes. 
With the estimated parameters in the proposed model, we characterized
the clustered SNP distributions obtained from different chromosomes,
including the X chromosome. Whereas the proposed model is valid
irrespective of the ploidy (i.e., either diploid or haploid), our
analysis of estimated parameters accounts for the characteristics,
such as the pronounced population structure and the low mutation
rate, specific to the X chromosome.  
\section{MATERIALS AND METHODS}
\subsection{Data}
\label{sdata}
We use the genome-wide Hapmap SNP data of Phase III, which consists of
1,440,616 SNPs in 1184 reference individuals from 11 global ancestry
groups of three continental regions. The data are publicly
available and can be downloaded at http://hapmap.ncbi.nlm.nih.gov/. 
To investigate the population-specific differences, we extracted SNPs
that are polymorphic within each of a single population. For each of
11 global populations, we analyzed SNP distributions on 22 autosomes
and the X chromosome.  
Generally, the number of SNPs depends on the number of samples, the
chromosome, and the genetic diversity of the population.  
Thus, the number of identified SNPs may fluctuate with populations and
chromosomes. 

Table \ref{sstat} shows the number of SNPs identified in each
population summed over all chromosomes and each chromosome averaged
over 11 global populations.  
The number of SNPs for each population in Table \ref{sstat} roughly
illustrates the regional difference in the genetic diversity. The
populations that originated from Africa (ASW, LWK, MKK, YRI) have a larger
number of SNPs than other continental regions, indicating a
higher genetic diversity. On the other hand, the population from Asia
(CHB, CHD, JPT) have a smaller number of SNPs and a lower genetic
diversity than others. 
\begin{table}[!t]
\begin{center}
\caption{The number $N$ of SNPs in each of single population (left two
  columns) and in each chromosome averaged over 11 populations (right
  four columns). The 
  population names are abbreviated and the full names can be found in
  Ref.~\cite{hap}. Note that the number of SNPs in each single
  population and the total number of SNPs of all chromosomes are less
  than 1,440,616 SNPs obtained from all global populations.}
{\begin{tabular}{|c|c||c|c|c|c|} 
\hline 
pop. name & $N$ & chr. & $N$ &  chr. & $N$  \\ \hline
ASW & 1,399,533 & 1 & 102,848 & 13 & 46,752  \\
CEU & 1,269,095 & 2 & 103,994 & 14 & 40,899  \\ 
CHB & 1,181,090 & 3 & 86,600  & 15 & 37,884  \\ 
CHD & 1,173,514 & 4 & 77,006 & 16 & 39,392 \\ 
GIH & 1,260,550 & 5 & 79,132 & 17 & 33,734  \\ 
JPT & 1,154,331 & 6 & 82,545 & 18 & 36,802 \\ 
LWK & 1,366,422 & 7 & 67,937 & 19 &  23,092 \\ 
MEX & 1,324,625 & 8 & 67,576 & 20 & 32,426 \\ 
MKK & 1,385,579 & 9 & 57,444 & 21 & 17,557 \\ 
TSI & 1,266,622 & 10 & 66,389 & 22 & 17,826 \\ 
YRI & 1,340,306 & 11 & 63,333 & X & 41,408 \\ 
    &           & 12 & 61,202 & total & 1,283,778 \\ \hline
\end{tabular}
}{}
\end{center}
\label{sstat}
\end{table}
The clustering property of SNPs can be represented by taking the
ordered locations of all SNPs and examining how proximate they are 
located. We quantify the proximity of SNPs in terms of the SNP space, 
which is defined as the number of nucleotides between two adjacent
SNPs in their ordered locations. Specifically, let $\ell_i$ be the
location, in the number of nucleotides counting from $5^{\prime}$ of
a sequence, of the $i$th SNP in a chromosome. Then, the $i$th SNP space
$s_{i}$ is defined as 
\begin{equation}
s_i \equiv \ell_{i+1}- \ell_{i}, ~~ i=1,2, \cdots~.
\label{def}
\end{equation}
\subsection{Random model}
\label{rsec}
\begin{figure}[tb]
\centerline{
\includegraphics{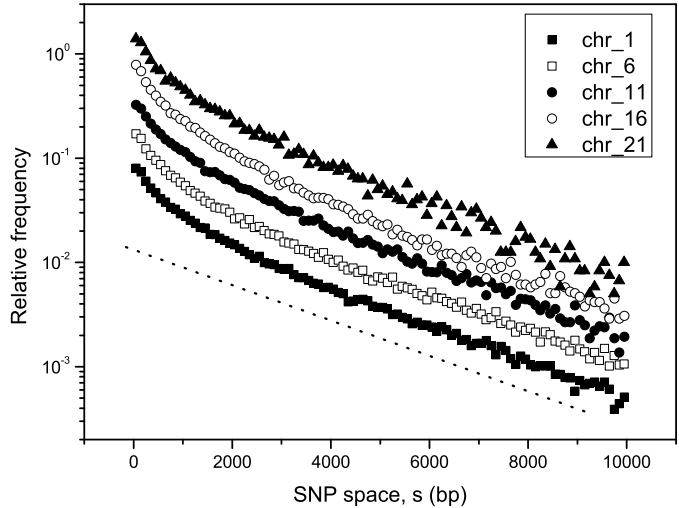}
}
\caption{Empirical frequency distributions of SNP space for different
  chromosomes of ASW. The dotted line represents a geometric
  distribution with an arbitrary slope. The vertical axis is in a log
  scale.}  
\label{fdist}
\end{figure}
An SNP is the result of a mutation and it is commonly assumed that the
mutation arises randomly in DNA sequence. Under this assumption, we
tested the hypothesis that SNPs are distributed randomly in a
sequence. In the random mutation model, mutations occur independently
of each other with a constant probability $\beta$ of $0 < \beta < 1$.  
With the random model, the probability distribution of the SNP space
defined in Eq.~(\ref{def}) can be derived as follows.
Suppose that an SNP is found at the location $\ell_k$. Then, the
probability of finding a subsequent SNP at the location
$\ell_{k+1}=\ell_{k}+s$ is given as    
\begin{equation}
p(s\,|\beta)=(1-\beta)^{(s-1)}\beta~, ~~ \mbox{for}~~ s=1,2, \cdots 
\label{pd}
\end{equation}
Note that Eq.~(\ref{pd}) is the probability mass function of a
geometric distribution~\cite{pitman}.
The geometric distribution is the discrete analogue of the exponential
distribution and has a property of being memoryless. The
distribution is often used for modeling the number of trials until the
first success, in our case, an SNP. 

By taking a logarithm on the both sides of Eq.~(\ref{pd}), we get
\begin{equation}
\ln p(s\,|\beta)=\ln (1-\beta)~ s+ \ln \frac{\beta}{1-\beta}~. 
\label{null}
\end{equation}
This illustrates that $\ln p(s\,|\beta)$ is linear in $s$ with $\ln
(1-\beta) < 0$ being the proportionality, or a slope. The parameter
$\beta$ can be estimated, for example, by the maximum likelihood
estimation (MLE). Thus, if mutations occur randomly, we expect that
$\ln p(s\,|\beta)$ should be a straight line in $s$ 
with a negative slope of $\ln (1-\beta)$. 

In Fig.~\ref{fdist}, we plot the empirical distribution of
the SNP spacing $s$ for different chromosomes of ASW. 
A comparison with the random model reveals that the empirical
distributions do not 
follow a geometric distribution (the dotted line), especially for
small values of $s$ in which the major deficiency of the model occurs. 
In particular, Fig. \ref{fdist} shows that the probability increases
sharply and non-linearly as $s$ becomes small which suggests that the SNPs are
clustered. This tendency of the distribution is more or less
independent of the chromosome. This reflects that SNPs are distributed not
randomly, but clustered. From these we can infer that the random
mutation hypothesis is inadequate to explain the clustering
property of SNPs.
\subsection{Proposed model}
The clustered SNPs imply that when a mutation occurs, another mutation
is  more likely to occur as they are closer in their
locations. This suggests that the mutation probability
is not independent of the location but dependent on how close the
mutations are in their locations.
This non-independent mutation can be modeled after the mutation
probability being inversely proportional to some power of the
separation in nucleotides between two consecutive mutations. Formally,
given that a mutation occurs at the location $\ell_{k}$, the
probability of the next mutation at $\ell_{k}+d$ can be expressed as 
\begin{equation}
r(d)=\frac{\lambda}{d^{\alpha}}
\label{alt}
\end{equation}
with $0 < \lambda <1$ and $\alpha \ge 0$. 

The proposed model contains two parameters, $\lambda$ and $\alpha$, to
be estimated. The $\lambda$ is the probability of two 
mutations occurring consecutively in genomic locations, and the $\alpha$ is
the strength of the mutational non-independence. 
The larger is $\alpha$, the smaller mutation probability as $d$ increases. 
In particular, when $\alpha=0$, the proposed model reduces to the
random model of Section \ref{rsec}. 

It should be noted that the proposed model is a phenomenological
model to account for the observed clustered SNP distributions, rather
than a model based on molecular mechanism. 
That is, the phenomenological model does not implement any biological
mechanism of mutation, such as the heterozygote instability. 
Rather, this model is based on the fact that if the mutation
does not occur randomly, there must be some dependence on the SNP space. 

With the proposed model, the probability of SNP space $s$
is given, for $s=2, 3, \cdots$, as 
\begin{eqnarray}
p(s\,|\lambda, \alpha)& =& \left(1-\frac{\lambda}{1^{\alpha}}\right)
\left(1-\frac{\lambda}{2^{\alpha}}\right) 
\cdots \left(1-\frac{\lambda}{(s-1)^{\alpha}}\right)
\frac{\lambda}{s^{\alpha}} \nonumber \\
& = & \frac{\lambda\prod_{k=1}^{s-1}(k^{\alpha}-\lambda)}{
  (s\, !)^{\alpha}}~. 
\label{pmodel}
\end{eqnarray}
Or, by taking a logarithm of Eq.~(\ref{pmodel}), we have
\begin{equation}
\ln p(s|\lambda, \alpha) = \sum_{k=1}^{s-1} \ln
(k^{\alpha}-\lambda)+\ln \lambda - \alpha \ln (s\,!)~.
\label{lmodel1}
\end{equation}
As the right hand side of Eq.~(\ref{lmodel1}) is intractable
analytically and even computationally, we resort to an approximate
expression.    

It is known that there is one SNP in roughly every 1,000
base pairs. Thus, we assume $\lambda \ll 1$ and expand 
$\ln (k^{\alpha}-\lambda)$ around $\lambda =0$ to the first order in
$\lambda$ to get
\begin{equation}
\ln (k^{\alpha}-\lambda) \approx \alpha \ln k - k^{-\alpha} \lambda ~.
\end{equation}
Using an integral approximation of
\begin{equation}
\sum_{k=1}^{s-1} k^{-\alpha} \approx \int_{1}^{s} k^{-\alpha} dk =
\frac{s^{1-\alpha}-1}{1-\alpha} ~,
\end{equation}
and the Stirling's formula of $\ln (s\,!) \approx s \ln (s)
-s$, Eq.~(\ref{lmodel1}) can be expressed approximately as 
\begin{equation}
\ln p(s|\lambda, \alpha) \approx \frac{\lambda}{1-\alpha} 
\left( 1-s^{1-\alpha}\right) + \ln \lambda -\alpha \ln s ~. 
\label{prop1}
\end{equation}
To show that the probability of $s$ in the random model [Eq.~(\ref{null})]
is a special case of Eq.~(\ref{prop1}), we expand Eq.~(\ref{null}) to the
first order in $\beta$ to get 
\begin{equation}
\ln p(s|\beta) \approx \beta (1-s) + \ln \beta~.
\label{null1}
\end{equation}
We can easily show that Eq.~(\ref{prop1}) becomes Eq.~(\ref{null1}) when
$\alpha=0$.

The parameters $\lambda$ and $\alpha$ in the proposed model can be
estimated, for example, by MLE. Under the assumption that empirical
SNP space data $\vec{s}=\{s_{1}, s_{2}, \cdots , s_{n}\}$ are
independent and identically distributed samples, we consider the 
log-likelihood ${L}(\vec{s}\,|\lambda, \alpha)$. It is given as 
\begin{equation}
{L}(\vec{s}\, | \lambda, \alpha) = \frac{\lambda}{1-\alpha} \left\{ n-
  \sum_{i=1}^{n}s_{i}^{1-\alpha} \right\} + n \ln \lambda  - \alpha
  \sum_{i=1}^{n} \ln s_{i}~. 
\end{equation}
With the log-likelihood and MLE, we can estimate the parameters of $\lambda$
and $\alpha$ by solving two equations,
\begin{equation}
\left. \frac{\partial L}{\partial \lambda}\right|_{\hat{\lambda}} = 0
~~\mbox{and}~~ \left. \frac{\partial L}{\partial
    \alpha}\right|_{\hat{\alpha}} = 0~, 
\label{mle}
\end{equation}
simultaneously using, for instance, an iterative method~\cite{kelley}. 
\begin{figure}[tb]
\centerline{
\includegraphics{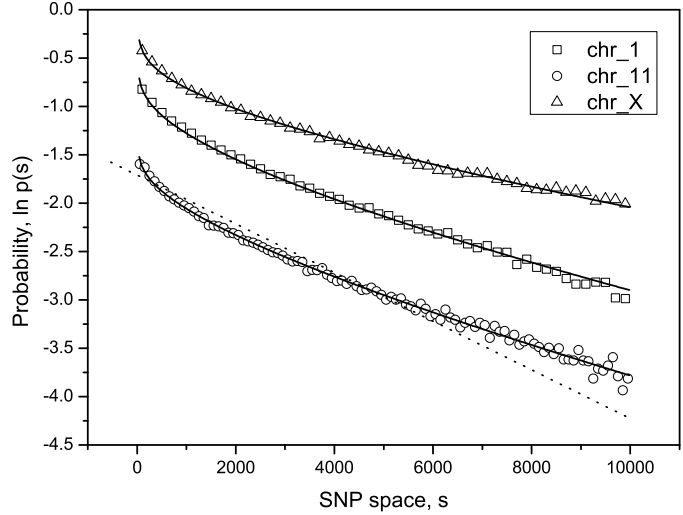}
}
\caption{Plots of empirical distributions of SNP space by using SNPs
  on ASW for the chromosome 1 (boxes), the chromosome 11 (circles),
  and the chromosome X (triangles), together with their distributions
  obtained by the proposed model (solid lines). The dotted line represents the
  distribution of the chromosome 11 obtained by the random model with
  the parameter $\beta$ estimated by MLE. The distributions of the chromosomes
  11 and X are shifted vertically for the display purpose, and the
  vertical axis is in a log scale.} 
\label{model}
\end{figure}

Although the phenomenological model that we used in this study is much
simpler and unrealistic than the heterozygote instability model, the 
phenomenological model is intractable mathematically unless we
make approximations. Thus, more realistic model such as the heterozygote
instability model is not easy to implement mathematically and/or analytically.
This is one of reasons why any biologically plausible
process hardly be modeled mathematically.
\section{RESULTS AND DISCUSSION}
\begin{figure}[tb]
\centerline{
\includegraphics{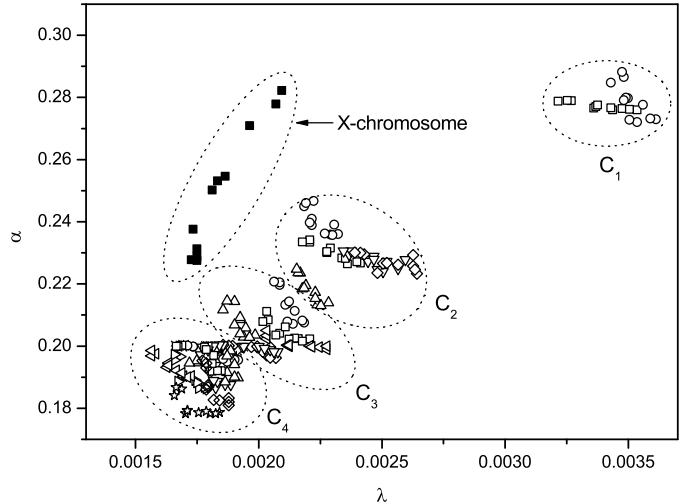}
}
\caption{Plots of estimated $\alpha$ versus $\lambda$ for all 253
  chromosomes (23 chromosomes from each of 11 global populations). The
  values for the X chromosomes from 11 global populations are plotted
  by filled boxes. The pairs of ($\alpha$, $\lambda$) for 22 autosomes
  are clustered into four clusters. They are:  $C_{1}=\{ 9(\Box),
  16(\bigcirc)\}$, $C_{2}=\{1(\Box), 15(\bigtriangleup),
  19(\bigtriangledown), 21(\bigcirc), 22(\Diamond)\}$, 
  $C_{3}=\{6(\Diamond), 7(\bigtriangleup), 8(\bigcirc), 10(\Box),
  11(\bigtriangledown), 20(\triangleleft)\}$, and  $C_{4}=\{ 2(\ast),
  3(\triangleright), 4(\triangleleft), 5(\bigcirc),
  12(\bigtriangledown), 13(\star), 14(\bigtriangleup), 17(\Box),
  18(\Diamond)\}$.}  
\label{all}
\end{figure}
\begin{figure}[tb]
\centerline{
\includegraphics{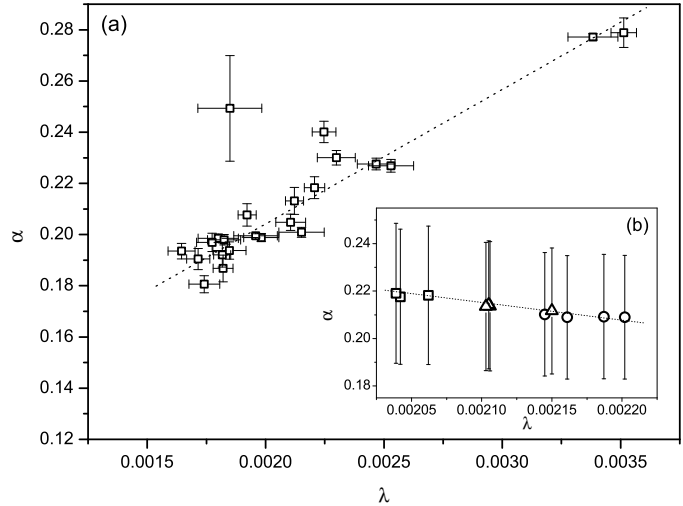}
}
\caption{(a) Plot of estimated $\alpha$ versus $\lambda$ for all
  chromosomes averaged over 11 global 
  populations. The vertical and horizontal error bars are the standard
  deviations of $\alpha$ and $\lambda$, respectively. (b) Plot of estimated
  $\alpha$ versus $\lambda$ averaged over 23 chromosomes for each single
  population: African populations (circles), Asian populations
  (boxes), and the other populations (triangles). The vertical
  error bars are the standard deviations of $\alpha$ and the standard
  deviations of $\lambda$ are omitted as they are too large to
  display. The dotted lines represent linear fits with the least
  square error.}  
\label{a1}
\end{figure}
Using Hapmap SNP data, we estimate two parameters $\lambda$
and $\alpha$ for all chromosomes of each single population. With the
estimated parameters, we present a few $\ln p(s\, | \lambda, \alpha)$,
together with the corresponding empirical distributions in
Fig.~\ref{model}.       
As shown in Fig.~\ref{model}, the model distribution fits well to the
empirical distribution, especially to small $s$ of non-linearly
increasing region representing clustered SNPs. Moreover,
Fig.~\ref{model} demonstrates that the proposed model is equally
applicable not only to the autosomes but to the X chromosome.  
We also see that the proposed model is more
adequate than the random model (dotted line in Fig.~\ref{model}).
This result supports, at least partially if not entirely,
the non-independent mutation hypothesis.

As the X chromosome is haploid in males, it has smaller population
size than the autosome has. Thus, we may expect that the clustering
property of SNPs on the X chromosome may differ from that on the
autosomes~\cite{amos}. A consequence of a smaller population size of
the X chromosome is a low abundance in SNPs.    
It is known that the nucleotide mutation rate in females is much lower than
in males~\cite{li}. The smaller population size and low mutation
rate result in the genetic diversity of the X chromosome about a 
half of that of the autosomes~\cite{schaff}. Thus, we expect that SNPs
on the X chromosome to be about half abundance that of
the autosomes~\cite{amos,schaff}. From the clustering perspective,
this implies that SNPs on the X chromosome are less clustered than
those on the autosomes in the sense that the average of SNP space $s$
over the X chromosome is larger than that over the autosomes.  
That is, larger $s$ is more probable on the X chromosome than on the
autosomes. This characteristic can also be seen by the visual
inspection of Fig.~\ref{model}.   

To measure the abundance in SNPs, we fit the empirical distribution of
$\log p(s)$ in terms of a linear function (i.e., the random model of
Eq.~(\ref{null})) in $s$ so that the slope $\ln (1-\beta)$ of
the fitted function approximately quantify the degree of clusteredness. 
We estimated the slope for all chromosomes, and found that the slopes
were in the range of $[-1.70 \times 10^{-4}, ~-2.24\times 10^{-4}]$
for the autosomes and $-1.42\times 10^{-4}$ for the X
chromosome. The fact that the slope for the X chromosome is less steep than
that for the autosomes quantitatively supports the low abundance and
fewer clusters of SNPs on the X chromosome. 

Based on the finding that the proposed model can explain
the clustering property of SNPs, we further analyze the estimated parameters
of the proposed model. Figure \ref{all} plots estimated $\alpha$
versus $\lambda$ using SNP data on 23 chromosomes from each of 11
global populations.  
From Fig.~\ref{all}, we see that the parameters for the X chromosome 
exhibits a larger variation across 11 global populations than those for the
autosomes. Being a haploid, the smaller population size of the X
chromosome has a faster genetic drift than the autosomes.
Consequently, the population structure (or population stratification)
of the X chromosome should be more pronounced than that of the
autosomes~\cite{schaff}. Given that the population structure implies
the presence of a systematic difference in allelic frequencies between
global populations, we expect that chromosomes having pronounced
population structure should have higher dissimilarity in their allele
between populations, which, in turn, induces higher variability in
clustered SNP distribution. Thus, the comparatively large variation 
of the X chromosome across the populations is evidence of its
pronounced population structure. 

We also clustered autosomes according to their estimated parameters of
$(\alpha, \lambda)$ by using the k-means algorithm~\cite{hart} with
the number of clusters $k=4$, suggested by the method for determining
$k$~\cite{charrad}. The result of cluster analysis of 22 autosomes are
shown in Fig.~\ref{all}. The cluster analysis result demonstrates that
chromosomes in the same cluster (for instance, chromosomes 9 and 16 in
$C_{1}$) have similar degree of clustered SNPs in terms of their probability
distribution of the SNP space.

From the finding that $\lambda \ll \alpha < 1$, we can further
approximate Eq.~(\ref{prop1}) to investigate the mutational non-independence. By
expanding Eq.~(\ref{prop1}) to the first order in $\alpha$, it can be
approximated as 
\begin{equation}
\ln p(s|\lambda, \alpha) \approx \lambda(1-s) + \ln \lambda +
(1-s+s\ln s)\alpha \lambda - \alpha \ln s~.
\label{appr1}
\end{equation}
One way to express the mutational non-independence part is the
difference between logarithmic probabilities of Eq.~(\ref{appr1}) and
Eq.~(\ref{null1}). By identifying $\beta = \lambda$, we have
\begin{equation}
\ln\Delta \equiv \ln p(s|\lambda, \alpha) - \ln p(s|\beta) = (\lambda -
\lambda s + \lambda  s \ln s)\alpha - \alpha \ln s ~. 
\label{app1}
\end{equation}
Note that the non-random part depends not only on the parameters
$\alpha$ and $\lambda$, but also the SNP space $s$. 
Since $\lambda \ll 1$ as we have seen in Fig.~3, we can approximate
the behavior of $\ln \Delta$ for both large and small values of $s$.
When $s \gg 1$, we have $\lambda s \approx O(1)$, and $\ln\Delta
\approx -\alpha$. This indicates that the non-random 
part is insensitive to $s$ for $s \gg 1$, which implies that the mutational
non-independence is negligible. Whereas, when  $s \approx O(1)$,
we have $\ln\Delta \approx - \alpha \ln s$. This suggests that as $s$
becomes smaller, the non-random part increases, which is consistent
with the results that can be found, for instance, in Fig.~\ref{fdist}.

To investigate the chromosome and population specific characteristics,
we represent the estimated parameters that are averaged over 11 global
populations for each chromosome in Fig.~\ref{a1} (a), and those that
are averaged over 23 chromosomes for each single population in
Fig.~\ref{a1} (b).  
We found that the estimated parameters have smaller variation
across the chromosomes [Fig.~\ref{a1} (a)] than across the populations
[Fig.~\ref{a1} (b)]. This implies that the clustering property depends
more on the chromosome than on the population. 
In particular, among the variations across chromosomes, Fig.~\ref{a1}
(a) shows that the chromosome X has higher variability than the
autosomes whose cause was discussed before.   

Aside from the deviations, we uncovered that the two parameters are
positively correlated over all autosomes, whereas they are negatively
correlated over the global populations.  
The mutation probability of Eq.~(\ref{alt}) increases (or decreases) as
$\lambda$ (or $\alpha$) increases. 
This implies that, when the parameters are positively correlated, the
increase in $\lambda$ is compensated for by the increase in $\alpha$, and
vice-verse. 
Thus, the positive correlation over autosomes demonstrates that the
mutation probability stays more or less the same for all autosomes. 

On the other hand, the negative correlation over the global populations
illustrates that the mutation rate depends on the estimated values of
$\lambda$ (or $\alpha$) of the population. That is, as $\lambda$
increases, a negatively correlated $\alpha$ decreases; as a result,
the mutation probability of Eq.~(\ref{alt}) increases.
Thus, the larger the $\lambda$ (or the smaller $\alpha$) with a single
population, the higher its mutation probability. It turns out that the
populations that originated from Africa 
have a larger $\lambda$ than those from Asia as indicated in
Fig.~\ref{a1} (b). 
\section{SUMMARY AND CONCLUSION}
In this study, we proposed a non-random mutation model to explain the
clustered distribution of SNPs in the human genome. The proposed model
takes into account the dependence of the mutation probability on the
space between two adjacent SNPs. The proposed model was tested against
Hapmap data by deriving the SNP space distribution from the model. The
probability distribution derived from the proposed model was in good
agreement with empirical distributions. Furthermore, the proposed model was
comprehensive in the sense that it explains the clustered distribution
of SNPs not only on the autosomes but on the X chromosome. 
This suggests that the observed SNP distributions of all chromosomes are 
different realization of the same model. 
We also showed that the proposed model was more adequate than the random
mutation model.  

The parameters in the proposed model are used to demonstrate the
chromosome- and population-wise different characteristics of the SNP
distribution. 
From the fact that the X chromosome is haploid in males, the X
chromosome has a more pronounced population structure and a lower
abundance in SNPs than the autosomes. These characteristics were
understood by a comparison of the parameters for the X chromosome
with those for the autosomes.  
The correlation between $\lambda$ and $\alpha$ can be used to
illustrate the dependence of the mutation probability and genetic
diversities on the different populations. 

This study introduces a phenomenological model for the clustered SNP
distribution in the human genome. 
Further investigations, such as the
feature of evolutionary transience of SNP clusters and phylogenetic
reconstruction, are desirable and should be carried out as a future
direction of study.   
\section*{Acknowledgments}
This work was supported by the Korea Research Foundation Grant funded
by the Korean Government (MOEHRD) (KRF-2015018708). 

\end{document}